\documentclass[apjl]{emulateapj}
\usepackage{psfig,amsfonts,amsmath,graphicx,natbib,apjfonts}
\def\har{1}
\def\prince{2}
\def\lbnl{3}

\shorttitle{Chromospheric Variability in SDSS M Dwarfs.  II.}
\shortauthors{Kruse et al.}

\begin{document}

\title{Chromospheric Variability in SDSS M Dwarfs.  II.
Short-Timescale H$\alpha$ Variability}

\author{
E.~A.~Kruse\altaffilmark{\har},
E.~Berger\altaffilmark{\har},
G.~R.~Knapp\altaffilmark{\prince},
J.~E.~Gunn\altaffilmark{\prince},
C.~P.~Loomis\altaffilmark{\prince},
R.~H.~Lupton\altaffilmark{\prince},
and D.~J.~Schlegel\altaffilmark{\lbnl}
}

\altaffiltext{\har}{Harvard-Smithsonian Center for Astrophysics, 60
Garden Street, Cambridge, MA 02138}

\altaffiltext{\prince}{Department of Astrophysical Sciences, Princeton
University, Peyton Hall, Ivy Lane, Princeton, NJ 08544}

\altaffiltext{\lbnl}{Lawrence Berkeley National Laboratory, 1 
Cyclotron Road, MS 50R5032, Berkeley, CA 94720}

\begin{abstract}
We present the first comprehensive study of short-timescale
chromospheric H$\alpha$ variability in M dwarfs using the individual
15 min spectroscopic exposures for $52,392$ objects from the Sloan
Digital Sky Survey.  Our sample contains about $10^3-10^4$ objects per
spectral type bin in the range M0--M9, with a total of about $206,000$
spectra and a typical number of 3 exposures per object (ranging up to
a maximum of $30$ exposures).  Using this extensive data set we find
that about $16\%$ of the sources exhibit H$\alpha$ emission in at
least one exposure, and of those about $45\%$ exhibit H$\alpha$
emission in all of the available exposures.  As in previous studies of
H$\alpha$ {\it activity} ($L_{\rm H\alpha}/L_{\rm bol}$) we find a
rapid increase in the fraction of active objects from M0--M6.
However, we find a subsequent decline in later spectral types that we
attribute to our use of a spectral type dependent equivalent width
threshold.  Similarly, we find saturated activity at a level of
$L_{\rm H\alpha}/L_{\rm bol}\approx 10^{-3.6}$ for spectral types
M0--M5, followed by a decline to about $10^{-4.3}$ in the range
M7--M9.  Within the sample of objects with H$\alpha$ emission, only
26\% are consistent with non-variable emission, independent of
spectral type.  The H$\alpha$ {\it variability}, quantified in terms
of the ratio of maximum to minimum H$\alpha$ equivalent width ($R_{\rm
EW}$), and the ratio of the standard deviation to the mean
($\sigma_{\rm EW}/\langle {\rm EW}\rangle$), exhibits a rapid rise
from M0 to M5, followed by a plateau and a possible decline in M9
objects.  In particular, $R_{\rm EW}$ increases from a median value of
about 1.8 for M0--M3 to about 2.5 for M7--M9, and variability with
$R_{\rm EW}\gtrsim 10$ is only observed in objects later than M5.  For
the combined sample we find that the $R_{\rm EW}$ values follow an
exponential distribution with $N(R_{\rm EW})\propto {\rm exp}[-(R_{\rm
EW}-1)/2]$; for M5--M9 objects the characteristic scale is $R_{\rm
EW}-1\approx 2.7$, indicative of stronger variability.  In addition,
we find that objects with persistent H$\alpha$ emission exhibit
smaller values of $R_{\rm EW}$ than those with intermittent H$\alpha$
emission.  Based on these results we conclude that H$\alpha$
variability in M dwarfs on timescales of 15 min to 1 hr increases with
later spectral type, and that the variability is larger for
intermittent sources.  Future studies using this large sample will
address the variability timescales, the variability of other
chromospheric emission lines (e.g., H$\beta$, \ion{Ca}{2} H\&K), and
the origin of the highest amplitude events.
\end{abstract}


\section{Introduction}

The study of magnetic activity in fully convective low mass stars and
brown dwarfs (spectral types late-M, L, and T) has progressed in
recent years from the question of whether these objects produce stable
fields to a quantitative investigation of how the fields are produced
and dissipated.  Unlike the solar-type $\alpha\Omega$ dynamo, which
operates in the transition region between the radiative and convective
zones (the tachocline; \citealt{par55}), objects later than spectral
type M3 can only support a convective dynamo.  Numerical simulations
of such a mechanism are still at an early stage, but they suggest that
large-scale axisymmetric fields can indeed be generated in fully
convective objects, at least for conditions that roughly correspond to
mid-M dwarfs ($M\sim 0.3$ M$_\odot$; \citealt{bro08}).  Thus,
observational constraints on the scale, geometry, and dissipation of
the fields are essential.

Several observational techniques are now being used to address these
questions, including Zeeman measurements in Stokes $I$ and $V$, which
probe the strength of the integrated surface fields and their
large-scale topology, respectively \citep{rb07,dmp+08,mdp+08}, and
activity indicators such as radio, X-ray, and H$\alpha$ emission,
which trace the dissipation of the field and hence its strength and
geometry (e.g., \citealt{whw+04,ber06,bbf+09}).  The activity
indicators also provide insight into magnetic heating, and their
temporal variability can potentially trace the field properties on
small scales that are inaccessible to the Zeeman measurements.  The
current results from Zeeman measurements point to a transition from
mainly toroidal and non-axisymmetric fields in M0--M3 dwarfs to
predominantly poloidal axisymmetric fields in mid-M dwarfs
\citep{dmp+08,mdp+08}, with field strength of $\sim 0.1-3$ kG
\citep{rb07,dmp+08,mdp+08}.  Studies of activity indicators point to a
rapid decline in X-ray and H$\alpha$ activity (i.e., $L_{X,{\rm
H\alpha}}/L_{\rm bol}$) at about spectral type M7, and uniform radio
luminosity (i.e., increasing $L_{\rm rad}/L_{\rm bol}$) at least to
spectral type L3 (\citealt{bbf+09} and references therein).  The
disparate trends may be due to a decoupling of the magnetic fields
from the increasingly neutral atmospheres, or to a shift in the
magnetic field configuration.

While X-ray and radio observations provide powerful insight into the
nature of the magnetic fields, H$\alpha$ chromospheric emission, which
traces gas at temperatures of $\sim 10^4$ K, is more easily accessible
for large samples as a by-product of standard optical spectroscopic
observations.  In recent years, large spectroscopic samples of M
dwarfs have become available through dedicated studies and large-scale
surveys such as the Sloan Digital Sky Survey (SDSS).  These extensive
samples have led to several important results concerning chromospheric
activity.  First, the fraction of objects that exhibit H$\alpha$
emission increases rapidly from $\sim 5\%$ in the K5--M3 dwarfs to a
peak of $\sim 80-100\%$ around spectral type M7, followed by a
subsequent decline to a few percent in the L dwarfs
\citep{gmr+00,whw+04,whb+08}.  Second, while the level of activity
increases with both rotation and youth in F--K stars, it reaches a
saturated value of $L_{\rm H\alpha}/L_{\rm bol}\approx 10^{-3.6}$ in
M0--M6 dwarfs, followed by a rapid decline to $L_{\rm H\alpha}/L_{\rm
bol}\approx 10^{-5}$ by spectral type L0 \citep{hgr96,gmr+00,whw+04},
and a breakdown of the rotation-activity relation \citep{mb02}.
Finally, a small fraction ($\lesssim 5\%$) of late-M and L dwarfs have
been serendipitously observed to exhibit H$\alpha$ flares that reach
the saturated emission levels found in the early-M dwarfs
\citep{lkc+03}.

To uniformly address the latter point --- H$\alpha$ variability --- we
recently carried out spectroscopic monitoring observations of about 40
M4--M8 dwarfs with known H$\alpha$ emission \citep{lbk09}.  With
observations of about 1 hr per source and a time resolution of $5-10$
min, we found that about $80\%$ of these objects exhibit H$\alpha$
variability on a wide range of timescales, ranging in amplitude from
tens of percent up to a factor of about 5.  Indeed, the timescale
distribution for variability ``events'' is nearly flat from 10 min to
1 hr, with fewer events on timescales below 10 min.  The variability
amplitudes follow an exponential distribution with a characteristic
scale of ${\rm Max(EW)/Min(EW)}-1\approx 0.7$.  Finally, we found
tentative evidence for increased variability with later spectral type.

Here, we extend our study of H$\alpha$ variability by three orders of
magnitude using SDSS time-resolved spectroscopic data for a sample of
52,392 M0--M9 dwarfs (Knapp et al. 2009 in preparation; hereafter
Paper I).  For the first time we take advantage of the individual 15
min exposures, with a typical number of 3 exposures per source.  Using
these data we can thus probe H$\alpha$ variability on timescale of
$15$ min to $\sim 1$ hr.  In this paper we focus on the H$\alpha$
variability amplitudes and their relation to the H$\alpha$ activity
level ($L_{\rm H\alpha}/L_{\rm bol}$) and spectral type.  We also
re-investigate the relation between H$\alpha$ activity and spectral
type using the individual 15 min spectra, rather than the SDSS
pipeline-combined spectra, which were used in previous work
\citep{whw+04,whb+08}.  In \S\ref{sec:obs} we define our sample
selection and describe the analysis method for measuring H$\alpha$
equivalent widths.  The results for H$\alpha$ variability and activity
are described in \S\ref{sec:results}.  Future papers will focus on the
variability timescales, the highest amplitude events, the variability
of other chromospheric emission lines (e.g., higher-order Balmer
lines, \ion{Ca}{2} H\&K, \ion{He}{1} lines), and the relation between
chromospheric variability and various source properties, such as age.

\section{Sample Selection and Data Analysis}
\label{sec:obs}

The details of the sample selection are described in Paper I, where
preliminary characteristics of the sample are also described.  Here we
provide a summary of our selection technique.  We examined the
combined spectra of all point sources in the SDSS DR7 database with
$z\lesssim 19.5$, $i-z>0.2$, and $r-i>0.5$ mag (uncorrected for
interstellar extinction).  Using a set of spectral templates,
including the cool star templates from \citet{bwh+07}, we determined
the best-fit spectral type for each object and removed all objects
with stellar spectra earlier than K, or with the spectra of galaxies
or quasars.  We further removed spectra determined to be deficient
according to the following criteria: inadequate signal-to-noise ratio;
poor sky subtraction; large wavelength regions with missing data; and
contamination by light from a nearby object or by emission from
diffuse interstellar ionized gas.  The resulting list contains 59,427
stars.  After removal of stars of spectral type K, L and T, and
removal of stars with subdwarf spectra, we find a total of 52,392 M
dwarf stars.  

\begin{figure}
\centerline{\psfig{file=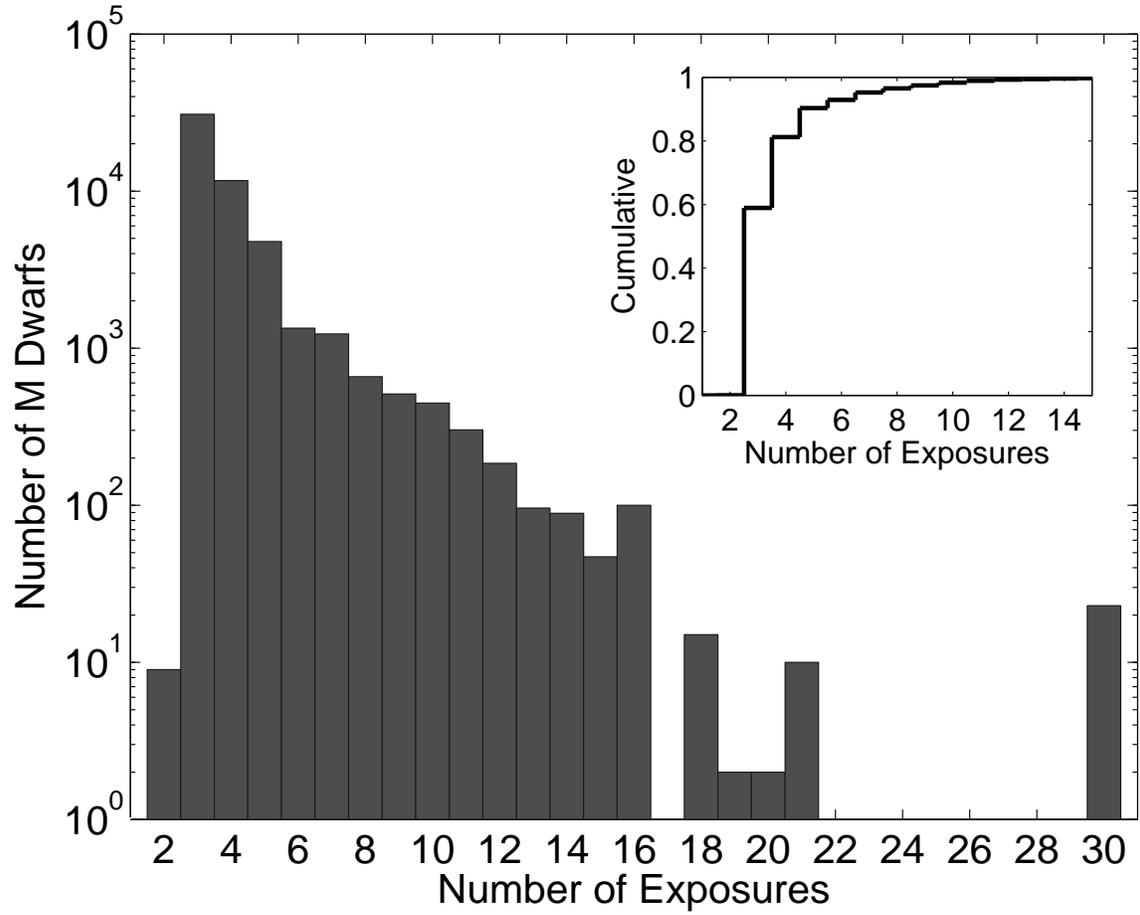,width=6.0in,angle=0}}
\caption{Histogram of the number of M dwarfs in our sample as a
function of the number of spectroscopic exposures.  About $60\%$ of
the objects have 3 exposures, $90\%$ have $3-5$ exposures, and $2.5\%$
have $\ge 10$ exposures.
\label{fig:number}}
\end{figure}

Using the coordinates of these objects, and their plate and fiber
identifications, we obtained the individual 15-min processed
spectroscopic exposures.  The total number of exposures is 205,823
with a range of $3-30$ exposures per source.  The distribution of the
number of exposures is shown in Figure~\ref{fig:number}.

\subsection{Equivalent Width Measurements}
\label{sec:ew}

\begin{figure}
\centerline{\psfig{file=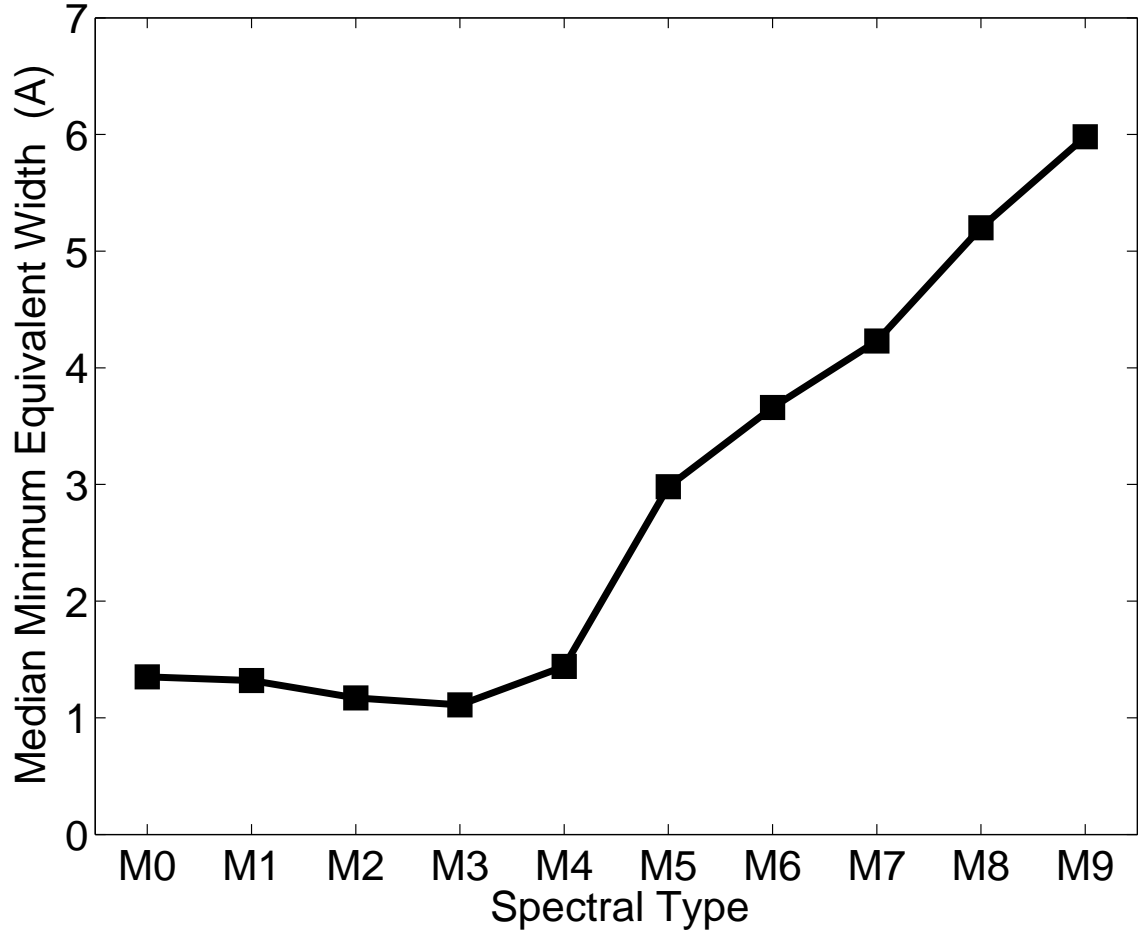,width=6.0in}}
\caption{Median value of the equivalent width $3\sigma$ limit as a
function of spectral type.  The rise in the equivalent width threshold
beyond M4 reflects the decrease in signal-to-noise ratio of the
spectra due to a reduction in the continuum flux level.
\label{fig:threshold}}
\end{figure}

To measure the H$\alpha$ equivalent width (EW) in each individual
spectrum we follow a three-step process.  First, we determine whether
H$\alpha$ emission is significantly detected.  We define the minimum
required H$\alpha$ emission line significance as $3\sigma$ relative to
the continuum level root-mean-square (rms) noise.  This condition
ensures that the H$\alpha$ emission line is significantly detected
regardless of the signal-to-noise ratio, and hence the apparent
brightness of the source (a combination of its spectral type and
distance).  In Figure~\ref{fig:threshold} we plot the median $3\sigma$
threshold as a function of spectral type.  For spectral types M0--M4
the threshold is about 1 \AA, but it then climbs rapidly to about 3
\AA\ for M6 and about 6 \AA\ for M9.  This spectral type dependent
threshold inevitably leads to different H$\alpha$ detection statistics
as a function of spectral type compared to previous analyses, which
used a fixed threshold of 1 \AA\ equivalent width from M0 to L0
\citep{whw+04,whb+08}; see \S\ref{sec:results}.  We note that since
the continuum flux in the H$\alpha$ region declines with later
spectral type, the use of a fixed equivalent width threshold allows
weaker, and less significant ``lines'' to be counted as genuine
detections.  Our spectral type dependent threshold therefore provides
more robust statistics on the H$\alpha$ properties at a fixed
statistical significance.  A total of 21,200 spectra exhibit H$\alpha$
emission at $>3\sigma$ significance ($10.3\%$ of all the available
exposures).

Second, we identify exposures with possible cosmic ray hits in the
H$\alpha$ line region using the standard SDSS pipeline flags.  We find
$1335$ exposures with such flags, or about $0.65\%$ of all spectra.
From visual inspection, however, we find that in about half of the
cases apparently genuine bright H$\alpha$ emission lines are flagged
as cosmic rays.  As a result, we perform a more robust cosmic ray
rejection using one of the two following criteria in addition to the
pipeline flag: (i) the flux of a flagged pixel within $\pm 2.5$ \AA\
of the H$\alpha$ line center is more than twice the maximum unflagged
pixel flux, or (ii) more than half of the pixels in the H$\alpha$ line
and continuum region ($6545-6595$ \AA) are flagged by the pipeline.
The former criterion allows us to distinguish cosmic rays with a
narrow width from genuine H$\alpha$ emission lines whose width is set
by the spectral resolution.  The latter criterion eliminates exposures
with questionable flux measurements in the H$\alpha$ line region.
Using these additional criteria we reduce the number of rejected
exposures to $746$ ($0.36\%$ of all exposures).

To evaluate the statistical efficacy of this procedure in removing
cosmic rays and in avoiding the removal of genuine H$\alpha$ emission
lines, we estimate the cosmic ray hit fraction in $5$ \AA\ windows
located in regions of the spectra that are devoid of emission
features.  Using a random set of about $13,000$ objects that cover the
full spectral type range of our study, we find that the expected
cosmic ray hit fraction in the H$\alpha$ line region is about
$0.30\%$, in good agreement with the fraction determined by our cosmic
ray rejection criteria.  This indicates that our sample is not biased
by cosmic rays, and furthermore that we are not missing a substantial
fraction of the large amplitude variability events due to cosmic ray
rejection.

Finally, to measure the H$\alpha$ line equivalent width (EW) we
simultaneously fit the H$\alpha$ line with a Gaussian profile and the
continuum region with a third-order Legendre polynomial.  The
uncertainty in the equivalent width ($\sigma_{\rm EW}$) is determined
over the same spectral region using the error spectrum associated with
each exposure.  For spectra that do not pass the $3\sigma$ threshold,
we use this value as an upper limit.  These upper limits are crucial
for assessing the range of activity for M dwarfs that exhibit a mix of
active and inactive spectra.

\begin{figure}
\centerline{\psfig{file=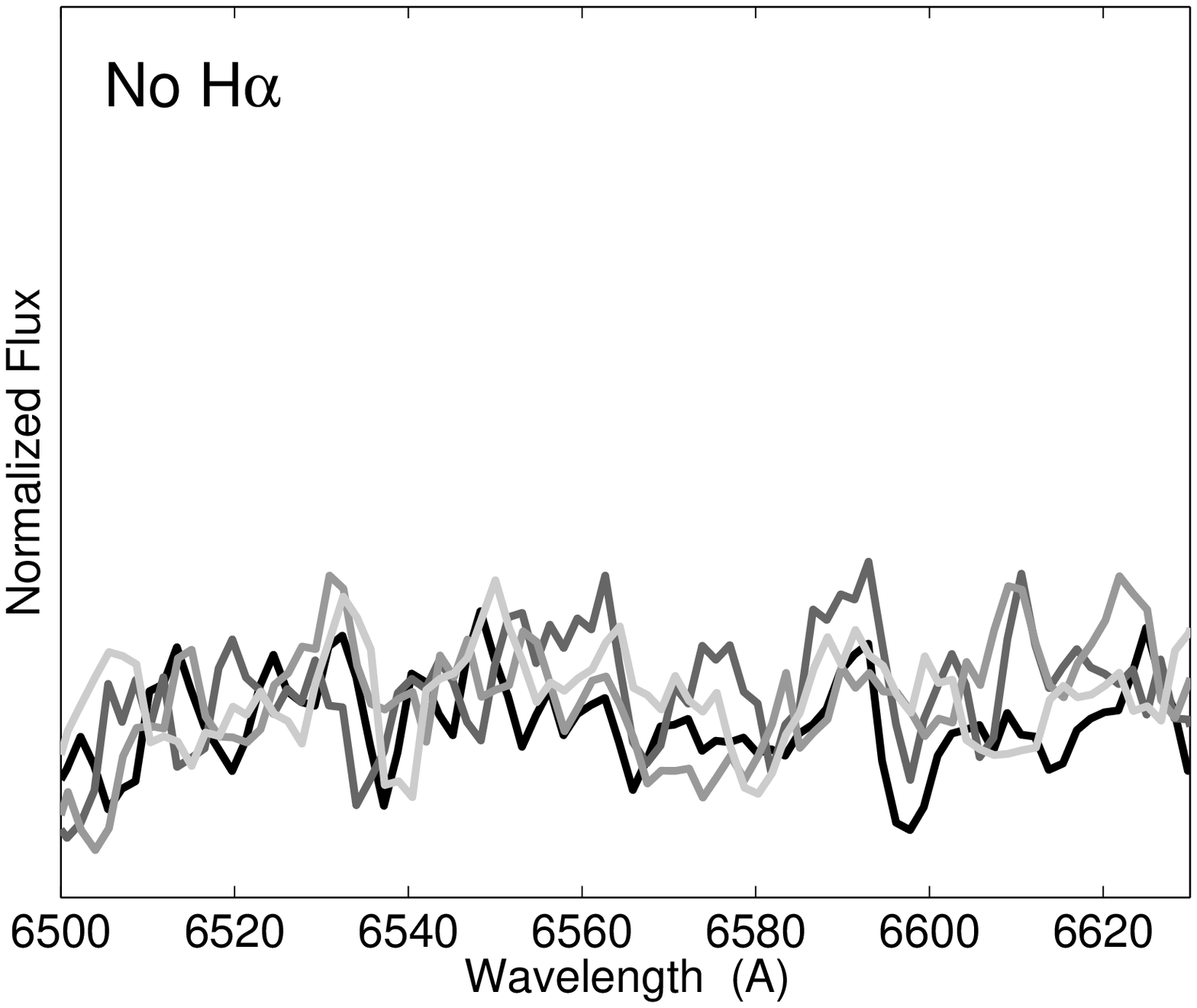,width=2.2in,angle=0}
\hspace{0.1in}\psfig{file=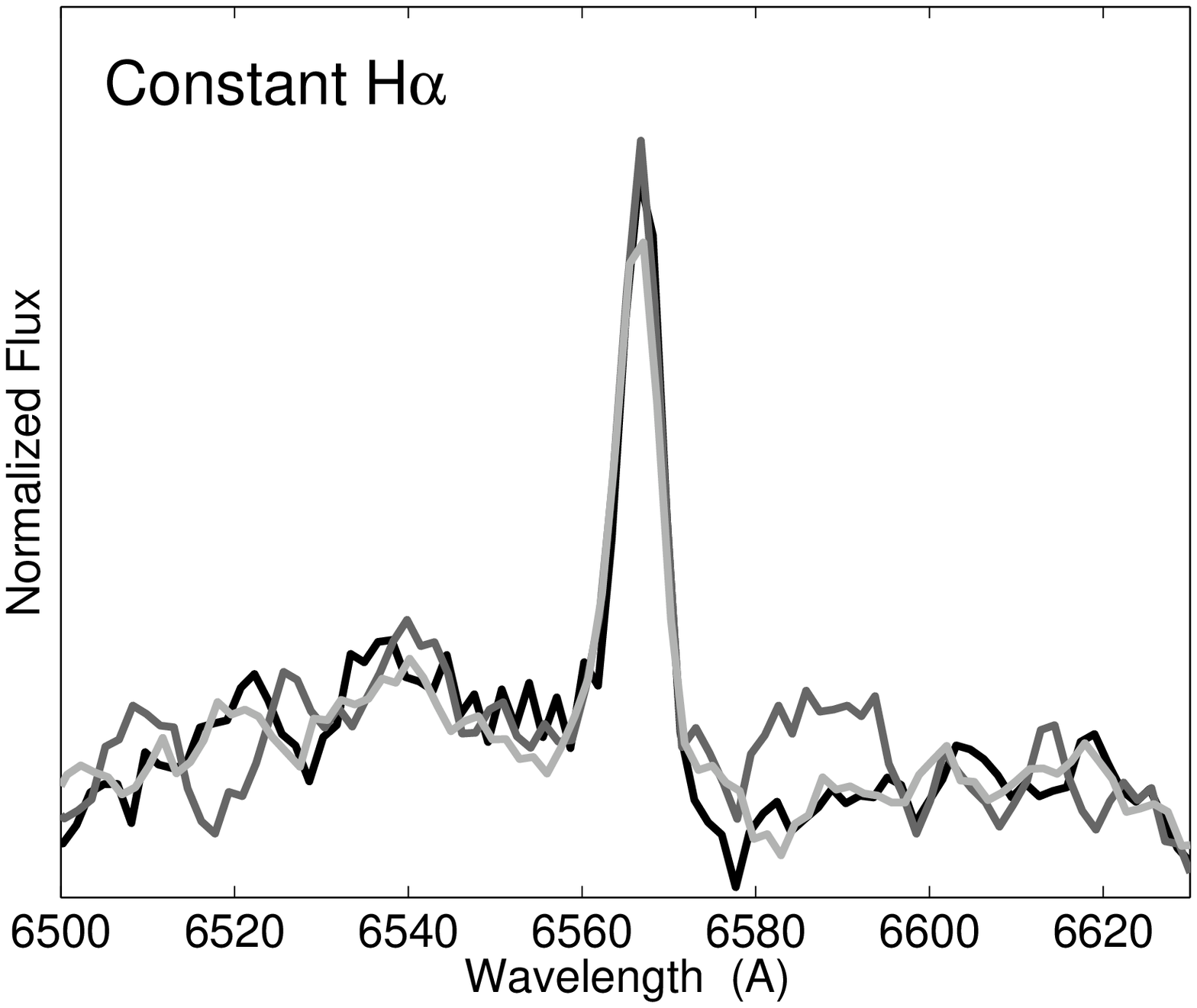,width=2.2in,angle=0}
\hspace{0.1in}\psfig{file=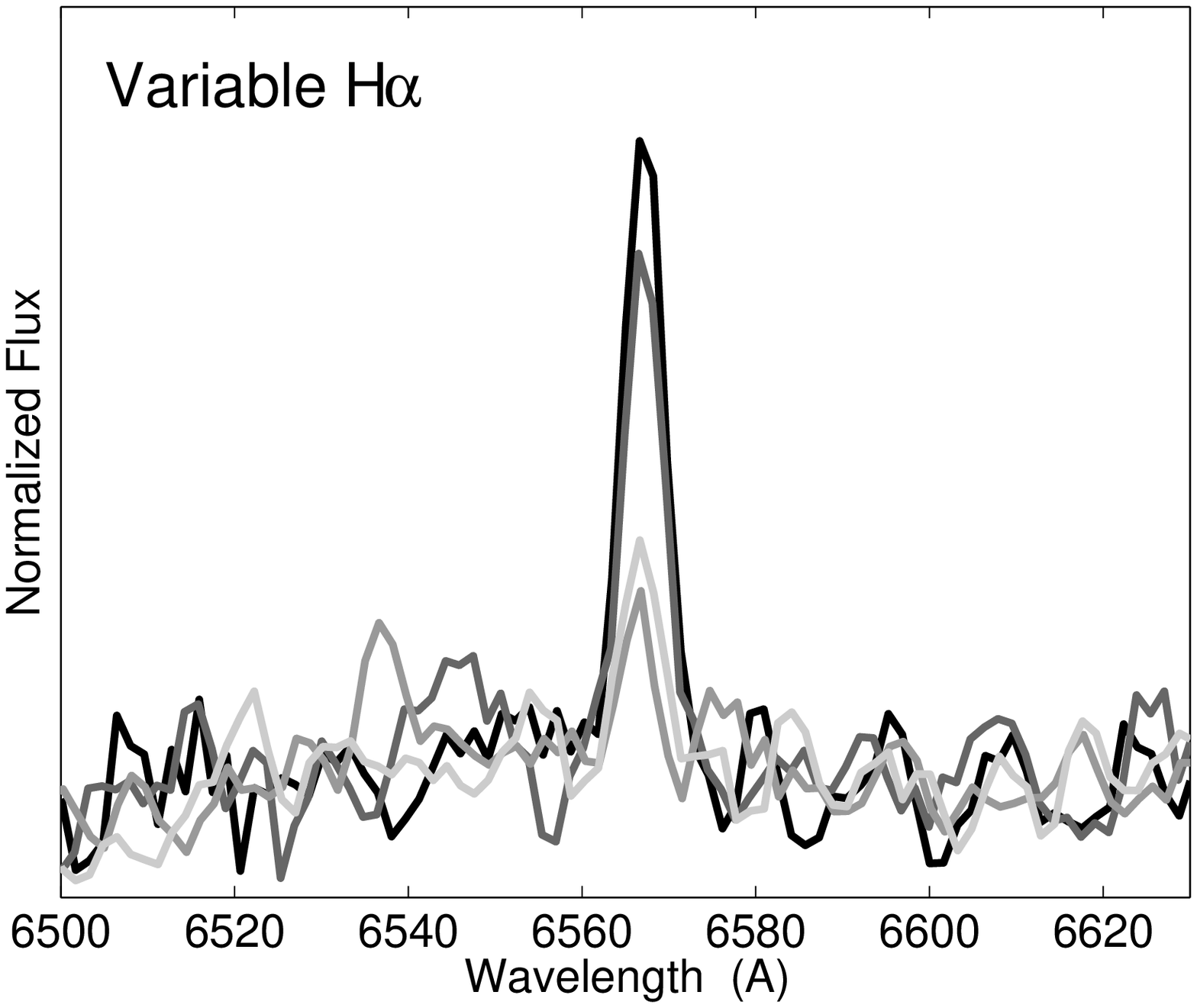,width=2.2in,angle=0}}
\caption{Representative examples of M dwarfs with no detectable
H$\alpha$ emission in any of the individual exposures (left), constant
H$\alpha$ emission (center) and variable H$\alpha$ emission (right).
In each panel the time sequence evolves from darker to lighter shade
of gray.
\label{fig:examples}}
\end{figure}

Example spectra with no H$\alpha$ emission, constant H$\alpha$
emission, and variable H$\alpha$ emission are shown in
Figure~\ref{fig:examples}.

\section{Results for H$\alpha$ Activity and Variability}
\label{sec:results}

\subsection{Detection Statistics}
\label{sec:stats}

\begin{figure}
\centerline{\psfig{file=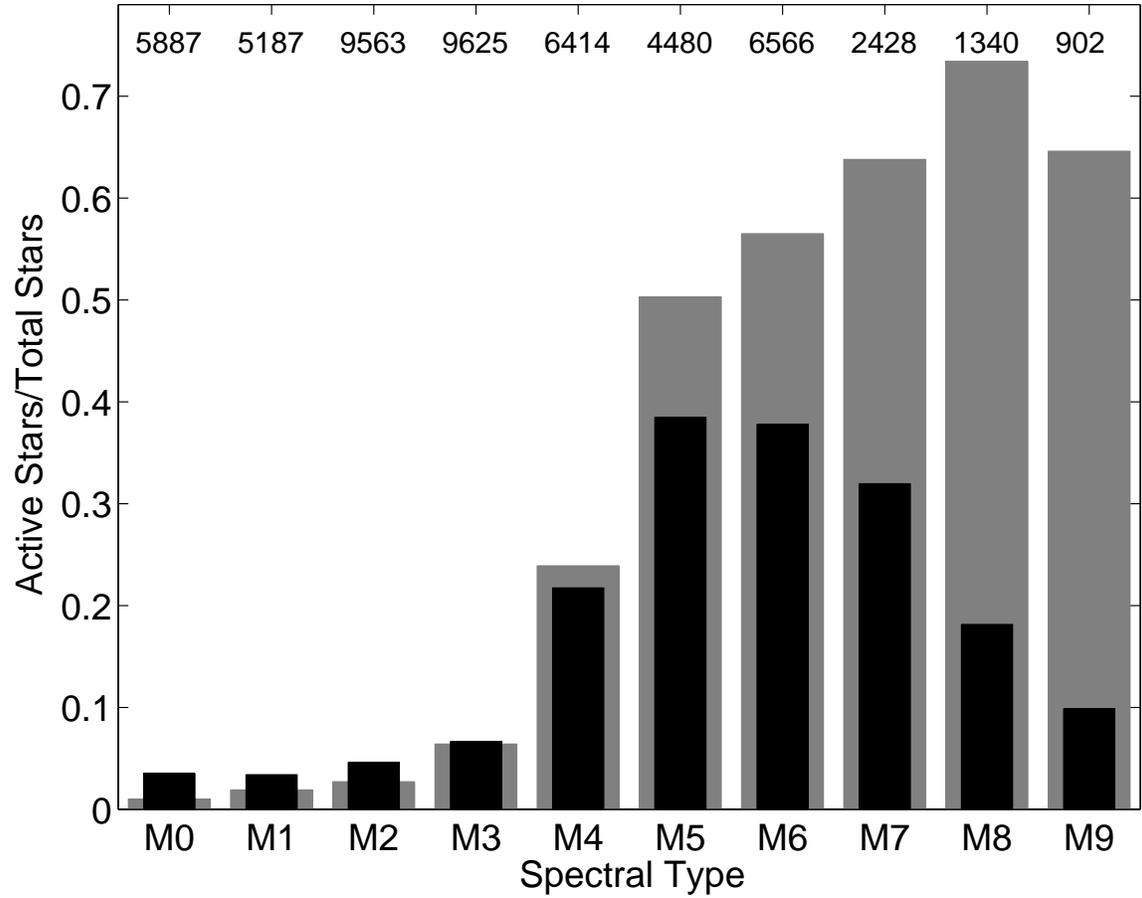,width=6.0in}}
\caption{Fraction of M dwarfs with at least one exposure exhibiting
detectable H$\alpha$ emission as a function of spectral type.  Our
results are shown in black, while the results from \citet{whw+04},
based on SDSS pipeline-combined spectra, are shown in gray.  The total
number of objects in each spectral type bin from our sample is shown
at the top.  The divergence between the two distributions beyond
spectral type M4 is primarily due to our use of a spectral type
dependent equivalent width threshold (Figure~\ref{fig:threshold}), as
opposed to a uniform threshold of 1 \AA\ from M0-L0 in \citet{whw+04};
see \S\ref{sec:stats}.
\label{fig:specdetections}}
\end{figure}

In our sample of 52,392 M dwarfs we find 8,223 objects ($15.7\%$) with
at least one spectroscopic exposure that exhibits H$\alpha$ emission.
This is verified by the independent analysis presented in Paper I,
which used the combined spectra.  The detection fraction as a function
of spectral type is shown in Figure~\ref{fig:specdetections}.  For
comparison, we also show the detection fractions from \citet{whw+04}
who used the pipeline-combined SDSS spectra and a fixed 1 \AA\
detection threshold.  As in previous studies, we find a rapid rise in
the H$\alpha$ detection fraction from a level of a few percent in the
range M0--M3 to about $20\%$ in M4.  However, beyond spectral type M4
our results diverge from previous studies.  Namely, we find that the
detection fraction rises to a peak of about $40\%$ in M5--M6, and
subsequently declines steadily to about $10\%$ in M9.  This is in
contrast to the continued rise from M4 to M8 (with a peak fraction of
nearly $100\%$) found by \citet{whb+08}.  This divergence is largely
due to our spectral type dependent detection threshold
(Figure~\ref{fig:threshold}).  We demonstrate that this is the case by
applying our threshold to the distributions of $L_{\rm H\alpha}/L_{\rm
bol}$ from \citet{whw+04}, using their equivalent width conversion
factors.  We find that the resulting reduction in active fraction is
about $15-30\%$, and it explains the difference for spectral types
M5--M7.  We still find a difference of about $20\%$ in the M8 and M9
bins.  It is possible that the continued rise found by \citet{whw+04}
and \citet{whb+08} was due to further contamination caused by their
use of a fixed equivalent width threshold.

\begin{figure}
\centerline{\psfig{file=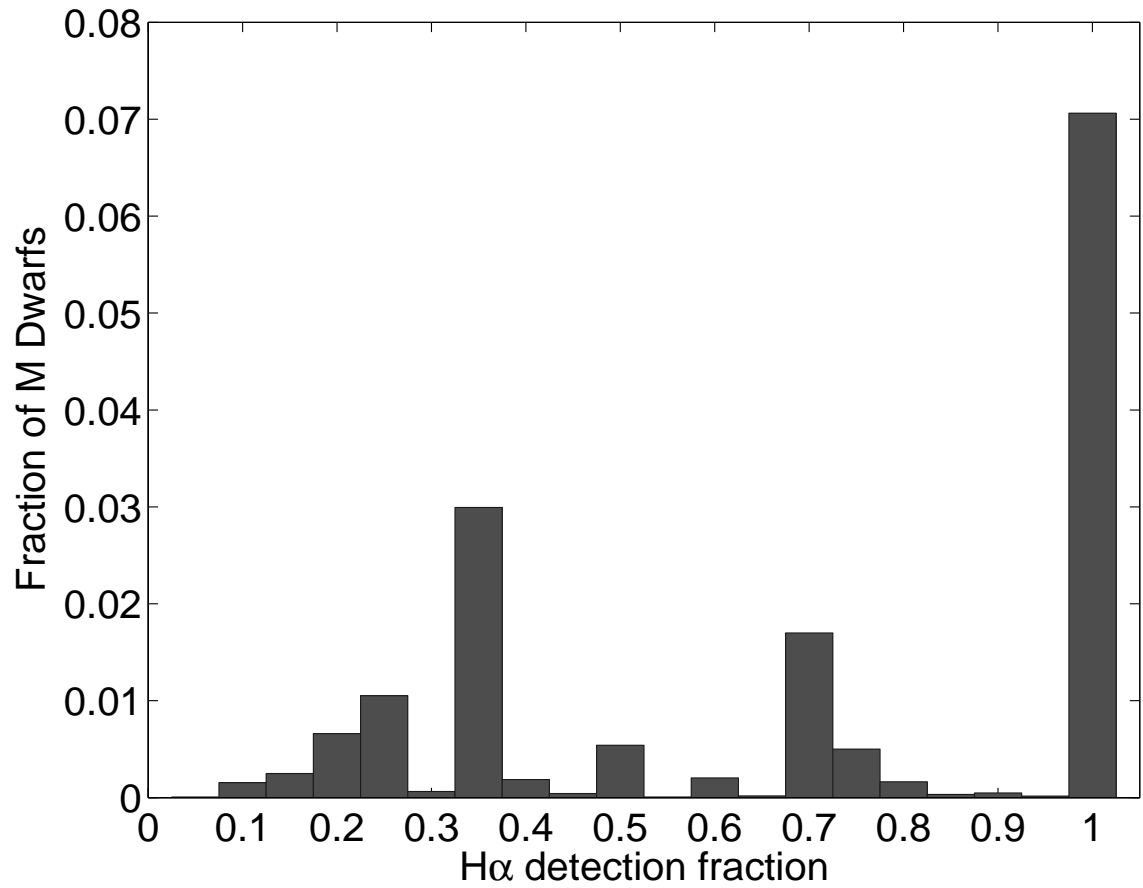,width=6.0in}}
\caption{Fraction of M dwarfs with detected H$\alpha$ emission binned
by the fraction of exposures in which H$\alpha$ emission is detected.
The peaks at 1/3, 2/3, and 1 are indicative of the large number of
objects with 3 exposures (Figure~\ref{fig:number}).  We do not show
the large spike at zero detection fraction, corresponding to about
$84\%$ of our sample.
\label{fig:detectionpercent}}
\end{figure}

Of the 8,223 objects with at least one active spectrum, $45\%$ exhibit
H$\alpha$ emission in all of the available exposures.  The fraction of
active objects as a function of the H$\alpha$ detection fraction is
shown in Figure~\ref{fig:detectionpercent}.  This distribution does
not take into account the fraction of objects in the sample with
various numbers of exposures (Figure~\ref{fig:number}).  For example,
the peaks at H$\alpha$ detection fractions of $1/3$, $2/3$, and $3/3$
reflect the large number of objects with 3 exposures.  Conversely, the
low fraction of objects with H$\alpha$ detection fractions of
$80-95\%$ simply reflects the small number of objects with a
sufficiently large number of exposures to populate these bins (i.e.,
more than 5 exposures).

Even without correcting for this effect we find that for the objects
with a small number of exposures ($3-5$), which occupy well defined
bins, there is a systematic decline in the fraction of objects as a
function of increased H$\alpha$ detection fraction.  For example, the
bins corresponding to $1/3$ and $2/3$ H$\alpha$ detection fractions
(dominated by objects with 3 exposures) exhibit a decline in the
fraction of active objects from $3.0\%$ to $1.7\%$.  Similarly, there
is a systematic decline in the fraction of stars with H$\alpha$
detections fractions of $1/4$, $2/4$, and $3/4$ (with values of
$1.1\%$, $0.5\%$, and $0.5\%$, respectively), and a decline for
objects with detection fractions of $1/5$ to $4/5$ from $0.7\%$ to
$0.2\%$.  We note that the rise at $100\%$ detection fraction partly
reflects the combined contribution from sources with a wide range of
exposure numbers, but also appears to be a genuine effect (some of the
objects in this bin have steady H$\alpha$ emission; see
Figure~\ref{fig:examples}).  The decrease in the fraction of objects
as a function of increased H$\alpha$ detection fraction may reflect a
typical timescale for the variability, such that variability on a
$\sim 15$ min timescale is more common than on $\sim 0.5-1$ hr
timescale.

\begin{figure}
\centerline{\psfig{file=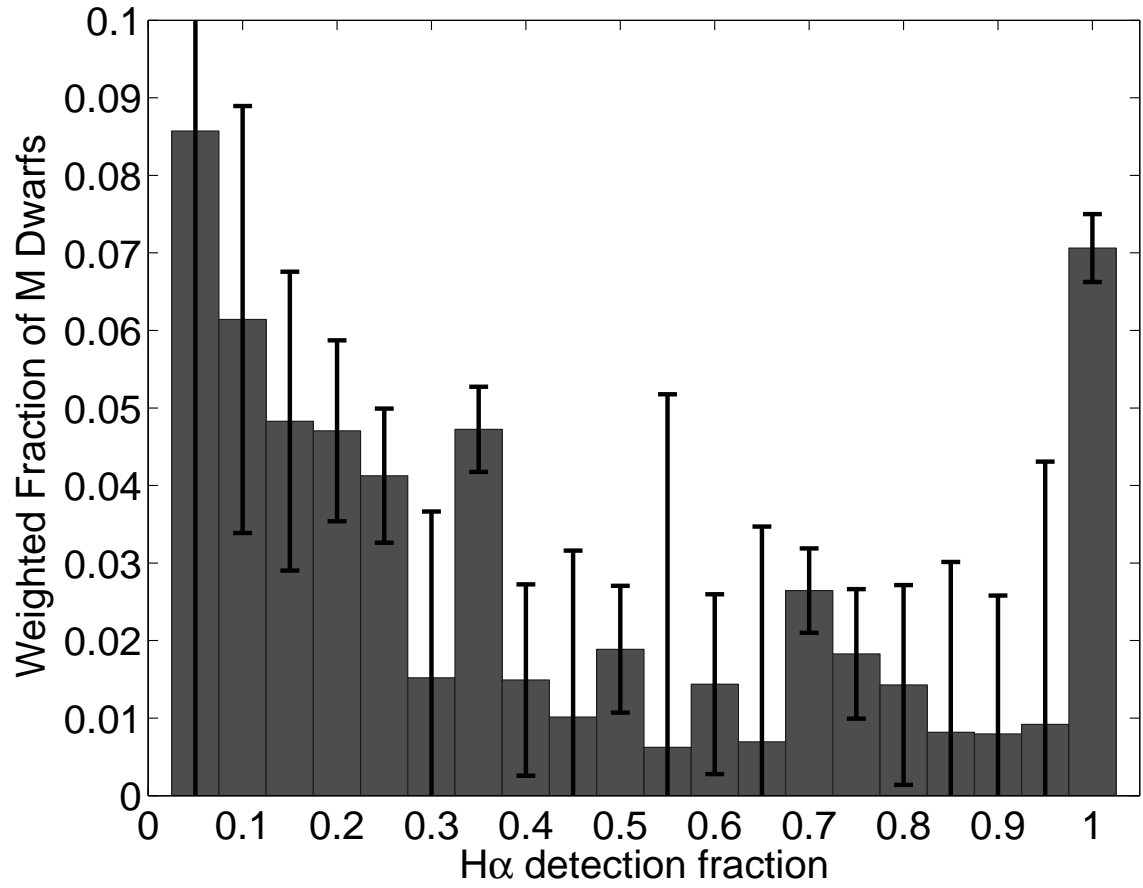,width=6.0in}}
\caption{Same as Figure~\ref{fig:detectionpercent} but corrected
appropriately for the number of M dwarfs in the sample that can
contribute to each H$\alpha$ detection fraction bin.
\label{fig:wdetectionpercent}}
\end{figure}

To appropriately normalize the distribution for the fraction of
objects with different numbers of exposures, we divide each H$\alpha$
detection fraction bin by the number of sources that can potentially
populate this bin.  For example, the bin at an H$\alpha$ detection
fraction of 0.5 is normalized by the total number of sources with an
even number of exposures, while the bin centered at 0.35 is normalized
by the number of objects with exposure numbers divisible by 3.  The
resulting normalized distribution is shown in
Figure~\ref{fig:wdetectionpercent}.  We find a relatively uniform
distribution below an H$\alpha$ detection fraction of about 1/3 at a
level of about $4\%$, followed by a decline to about $1-2\%$ for
detection fractions of $0.4-0.95$; the number of objects with an
H$\alpha$ detection fraction of 1 remains unchanged by definition.

These detection statistics alone already point to a significant level
of variability in M dwarfs.  Even if we make the most conservative
assumption that all of the objects with a $100\%$ H$\alpha$ detection
fraction are non-variable, the fraction of all M dwarfs with variable
H$\alpha$ emission is at least $8.6\%$.

\subsection{Activity}

\begin{figure}
\centerline{\psfig{file=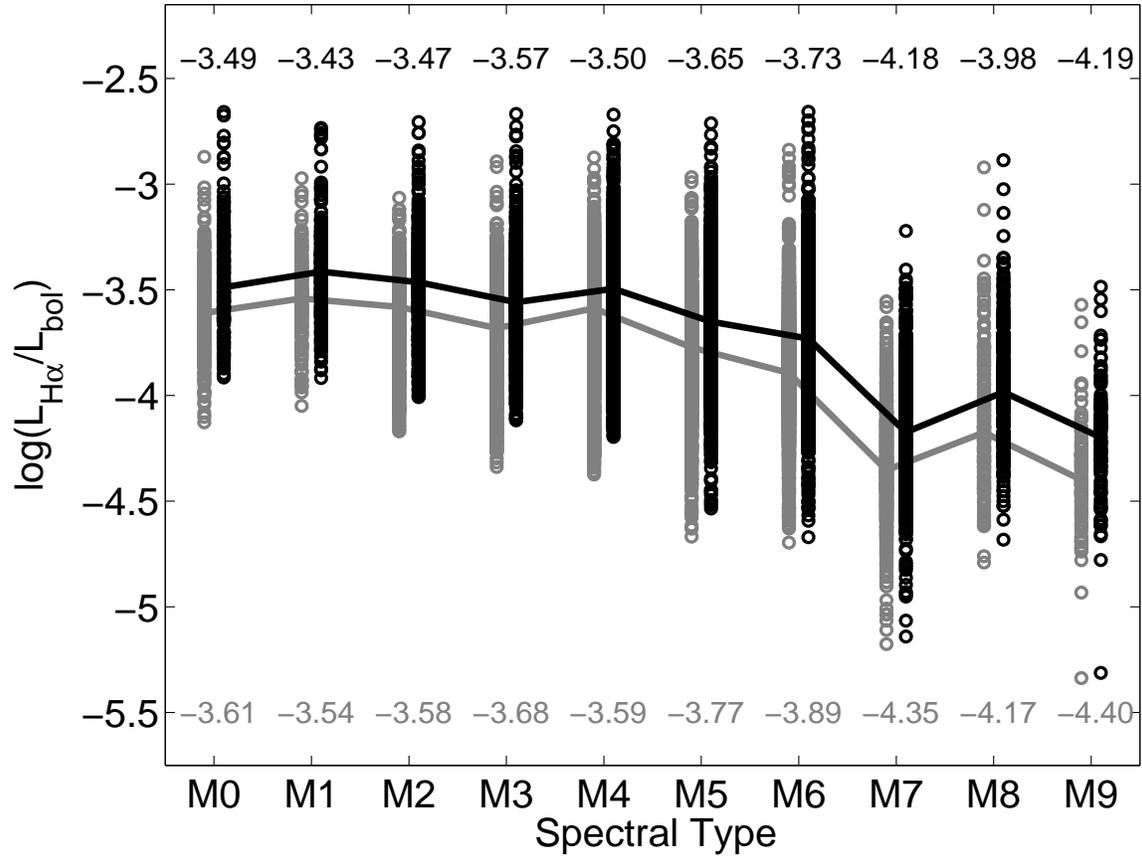,width=6.0in}}
\caption{H$\alpha$ activity level ($L_{\rm H\alpha}/L_{\rm bol}$)
plotted as a function of spectral type.  We use both the median (gray)
and the maximum (black) H$\alpha$ equivalent width for each object.
The lines connect the median value for each spectral type bin, and the
median values are listed at the top and bottom for the maximum and
median $L_{\rm H\alpha}/L_{\rm bol}$, respectively.
\label{fig:logbol}}
\end{figure}

Before we turn to a discussion of the H$\alpha$ {\it variability}, we
investigate the H$\alpha$ {\it activity} trends.  A comprehensive
analysis of M dwarf H$\alpha$ activity as a function of spectral type
based on SDSS pipeline-combined spectra was presented by
\citet{whw+04} and \citet{whb+08}.  In Figure~\ref{fig:logbol} we plot
$L_{\rm H\alpha}/ L_{\rm bol}$ as a function of spectral type for each
object.  We use both the median and the maximum values, and find
similar trends.  The median $L_{\rm H\alpha}/L_{\rm bol}$ value for
spectral types M0--M4 is roughly constant at $10^{-3.6}$, followed by
a decline to a value of about $10^{-4.3}$ in M7--M9.  These values are
similar to those found by \citet{whw+04}.  For the distribution of
maximum $L_{\rm H\alpha}/L_{\rm bol}$ we find a nearly constant value
$10^{-3.5}$ for M0--M4, followed by a decline to $10^{-4.1}$ in
M7--M9.  Thus, the maximum $L_{\rm H\alpha}/L_{\rm bol}$ values follow
the same trend as the median values, but the difference between the
median and maximum values appears to increase with later spectral
type, indicative of an increase in the H$\alpha$ variability.

\begin{figure}
\centerline{\psfig{file=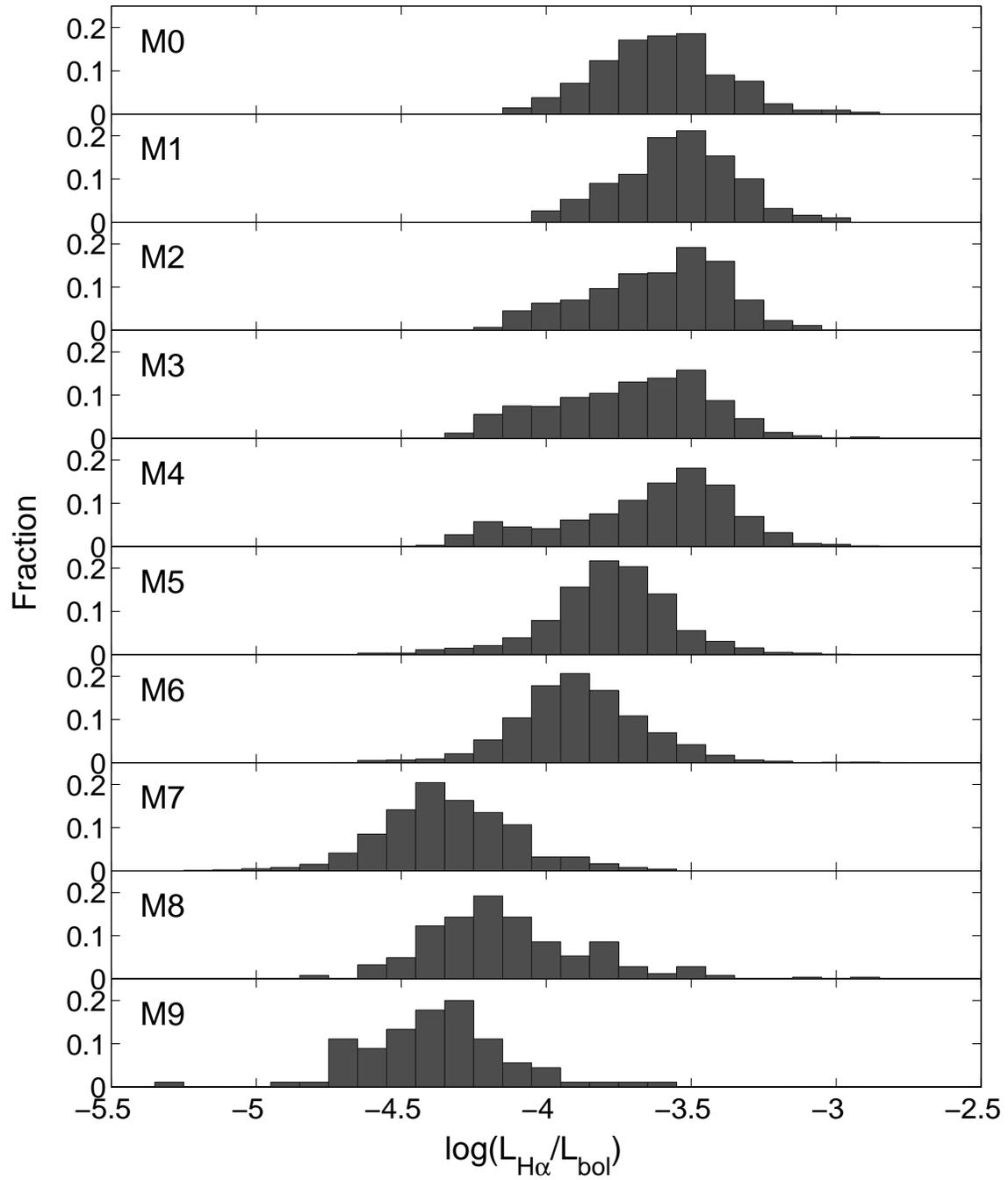,width=6.0in}}
\caption{Activity level ($L_{\rm H\alpha}/L_{\rm bol}$) histograms for
each spectral type bin.
\label{fig:logbolspectral}}
\end{figure}

In Figure~\ref{fig:logbolspectral} we plot the distribution of the
median $L_{\rm H\alpha}/L_{\rm bol}$ values in separate spectral type
bins.  The shift in the peak of the distribution to lower values as a
function of later spectral type is apparent, and follows the same
trend found by \citet{whw+04}.  However, unlike these authors we find
a wider dispersion in the early spectral types, primarily as a result
of the significantly larger number of objects in our sample.  The
standard deviation values range from about 0.2 to 0.28 with no clear
trend.

\subsection{Variability}

We now turn to the primary focus of our study --- H$\alpha$
variability.  We quantify the variability of each object using the
ratio of maximum to minimum H$\alpha$ equivalent width, $R_{\rm
EW}\equiv {\rm max(EW) /min(EW)}$, as well as the ratio of the
standard deviation to the mean equivalent width, $\tilde{\sigma}_{\rm
EW}\equiv\sigma_{\rm EW}/\langle {\rm EW}\rangle$.  In the case of
$R_{\rm EW}$ we calculate a lower limit on the variability when at
least one equivalent width upper limit is available, while for
$\tilde{\sigma}_{\rm EW}$ we treat the upper limits as $3\sigma$
detections; the resulting values are thus lower limits on the
variability.  Within the sample of objects with H$\alpha$ emission in
at least one spectrum (8,223 objects) we find that about $26\%$ are
consistent with steady H$\alpha$ emission within the uncertainties on
the individual H$\alpha$ equivalent width measurements.  This fraction
is relatively uniform across the full spectral type range of M0--M9,
with a possible mild decline in the later spectral types.

\begin{figure}
\centerline{\psfig{file=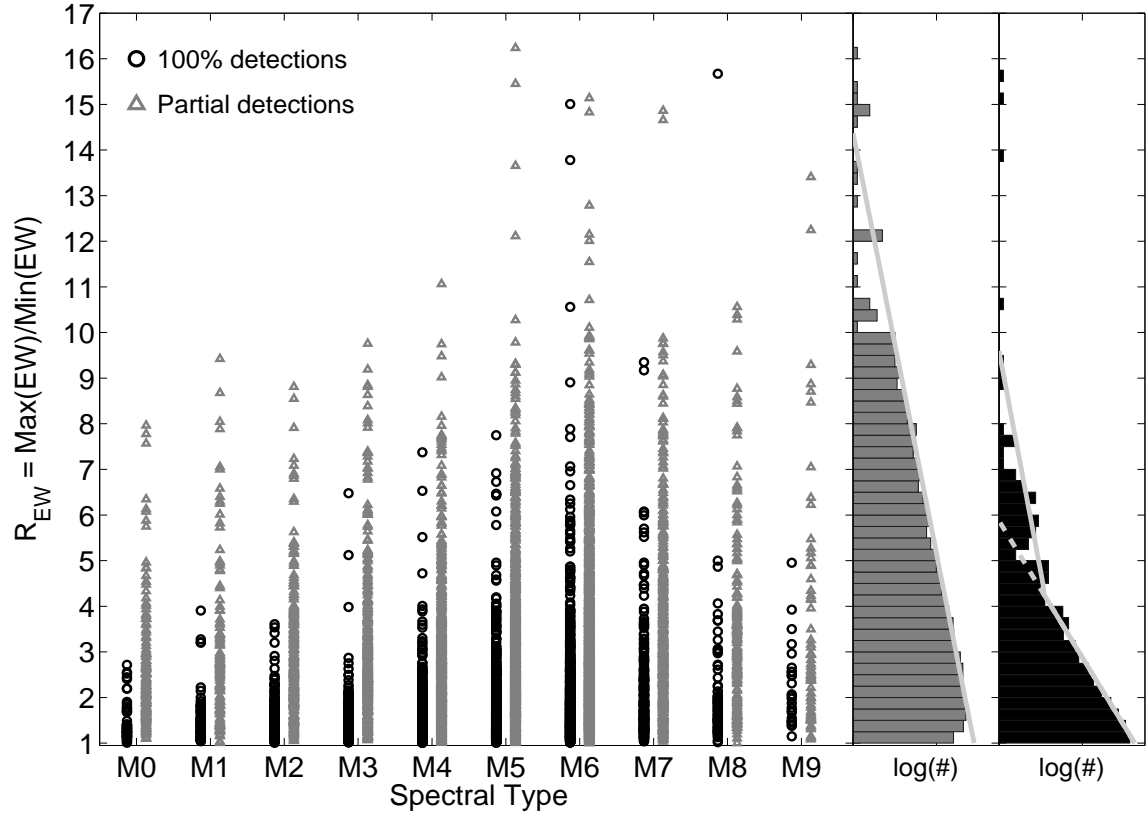,width=6.0in,angle=90}}
\caption{Ratio of maximum to minimum equivalent width ($R_{\rm EW}$)
for objects with detected H$\alpha$ emission as a function of spectral
type.  Circles designate objects classified as active in every
exposure, while triangles mark objects with at least one active
exposure and using upper limits on the minimum equivalent width for
the inactive spectra.  The histograms for both samples are shown in
the right panels.  The light gray lines are an exponential fit to
these samples.  The distribution for $100\%$ detection fraction
appears to exhibit a break at $R_{\rm EW}\approx 4.5$.
\label{fig:maxminlinear}}
\end{figure}

\begin{figure}
\centerline{\psfig{file=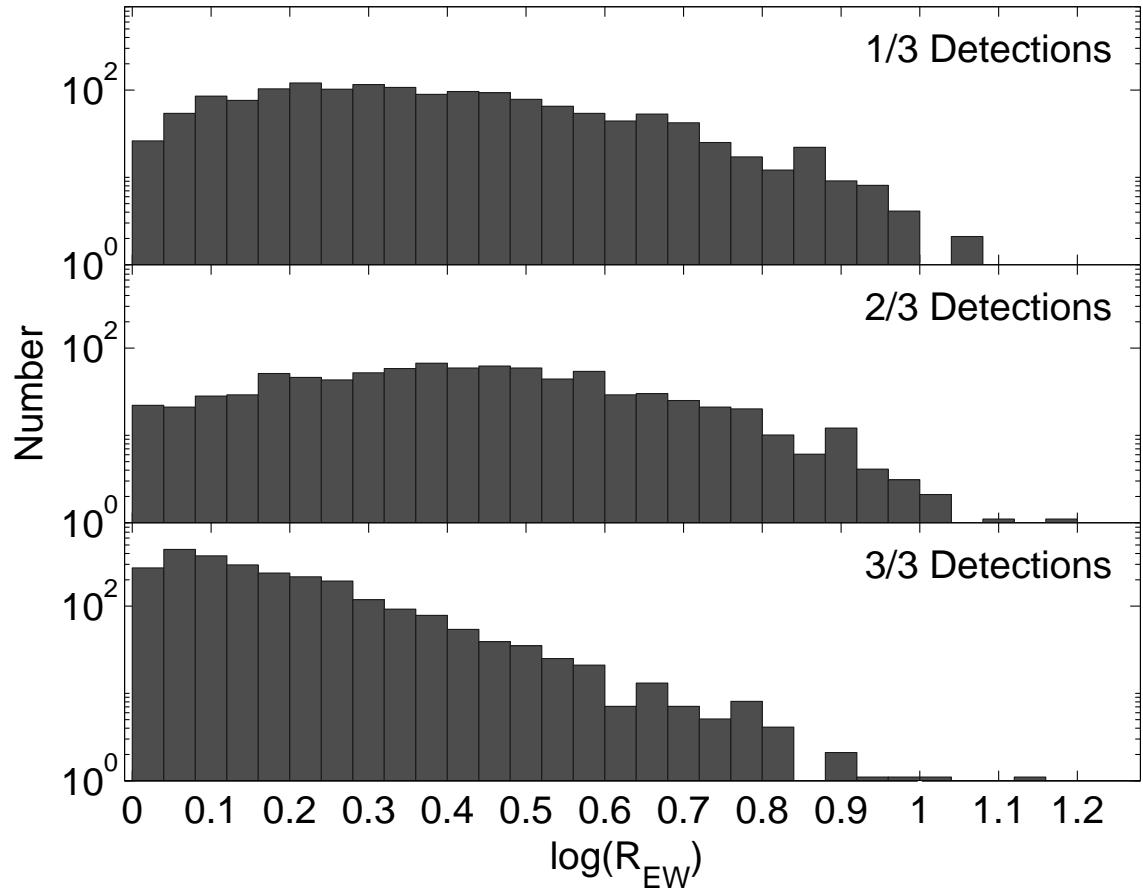,width=6.0in}}
\caption{Distribution of the ratio of maximum to minimum equivalent
width ($R_{\rm EW}$) binned by H$\alpha$ detection fraction for all
objects with 3 exposures.  The $R_{\rm EW}$ values for $1/3$ and $2/3$
detections are lower limits.  The objects with partial detections
exhibit higher amplitude variations.
\label{fig:3exposures}}
\end{figure}

\begin{figure}
\centerline{\psfig{file=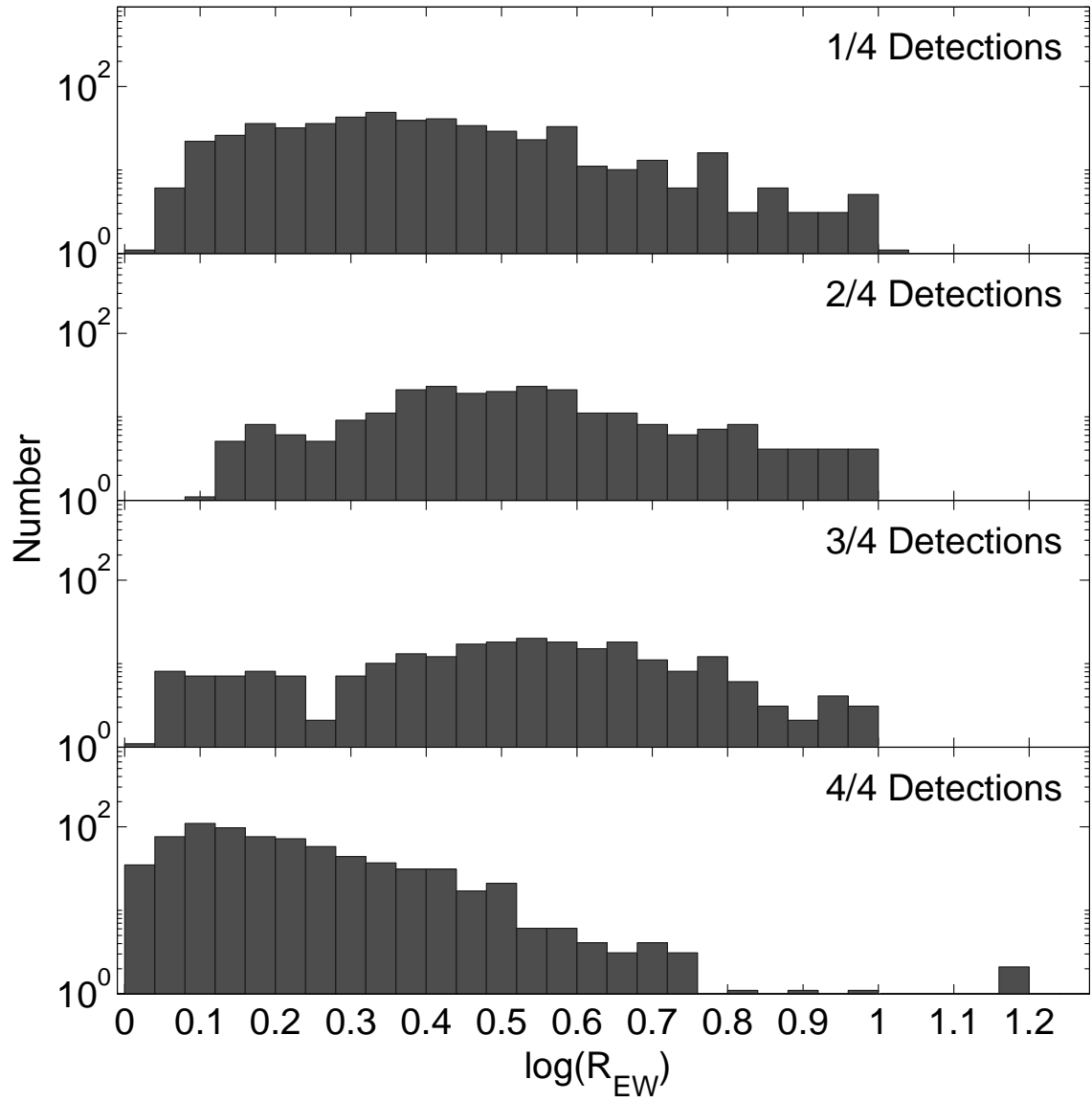,width=6.0in}}
\caption{Same as Figure~\ref{fig:3exposures} but for all objects with
4 exposures.
\label{fig:4exposures}}
\end{figure}

The distribution of $R_{\rm EW}$ values as a function of spectral type
is shown in Figure~\ref{fig:maxminlinear}.  For each spectral type bin
we separate objects with a $100\%$ detection fraction from those with
at least one upper limit.  In both cases we find a clear rise in
$R_{\rm EW}$ as a function of spectral type, from a maximum value of
about 5 for M0--M3 to about $10-17$ for M5--M8, followed by a possible
decline in spectral type M9.  Indeed, the objects that exhibit an
H$\alpha$ variability of $\gtrsim 10$ are nearly all M5--M9 dwarfs.
For the objects with a $100\%$ detection fraction we find that $1.2\%$
have $R_{\rm EW}>5$, while $0.1\%$ have $R_{\rm EW}>10$.  For the
objects with partial detections the fractions are $13.2\%$ and
$0.5\%$, respectively.  These fractions, and the overall wider
distribution of $R_{\rm EW}$ values for objects with partial
detections, indicate that objects with intermittent H$\alpha$ emission
produce higher variability ratios than those with persistent H$\alpha$
emission.  To investigate this point in more detail, we plot the
distributions of $R_{\rm EW}$ values binned by detection fraction for
all objects with 3 and 4 exposures (Figures~\ref{fig:3exposures} and
\ref{fig:4exposures}).  In both cases we find that the distributions
for objects with partial detections are significantly broader than for
objects with $3/3$ and $4/4$ detections.

We further find that the $R_{\rm EW}$ values follows an exponential
distribution (Figure~\ref{fig:maxminlinear}).  For the objects with a
$100\%$ detection fraction we find $N(R_{\rm EW})\propto{\rm exp}
[-(R_{\rm EW}-1)/0.7]$ for $R_{\rm EW}\lesssim 4.5$.  This result is
identical to what we previously found for a sample of about 40
objects, which reached a maximum value of $R_{\rm EW}\approx 5$
\citep{lbk09}.  The \citet{lbk09} sample was too small to provide
constraints on larger variability ratios.  Here we find that for
higher ratios ($R_{\rm EW}\gtrsim 4.5$), the distribution appears to
flatten to an exponential profile with $N(R_{\rm EW})\propto{\rm
exp}[-(R_{\rm EW}-1)/2]$.  The distribution of $R_{\rm EW}$ lower
limits is generally flatter, and appears to follow a single
exponential profile with $N(R_{\rm EW})\propto{\rm exp}[-(R_{\rm
EW}-1)/2]$.  Since these are lower limits, we conclude that the
overall distribution of $R_{\rm EW}$ values for M dwarfs is an
exponential with a characteristic scale of $R_{\rm EW}\gtrsim 2$.

\begin{figure}
\centerline{\psfig{file=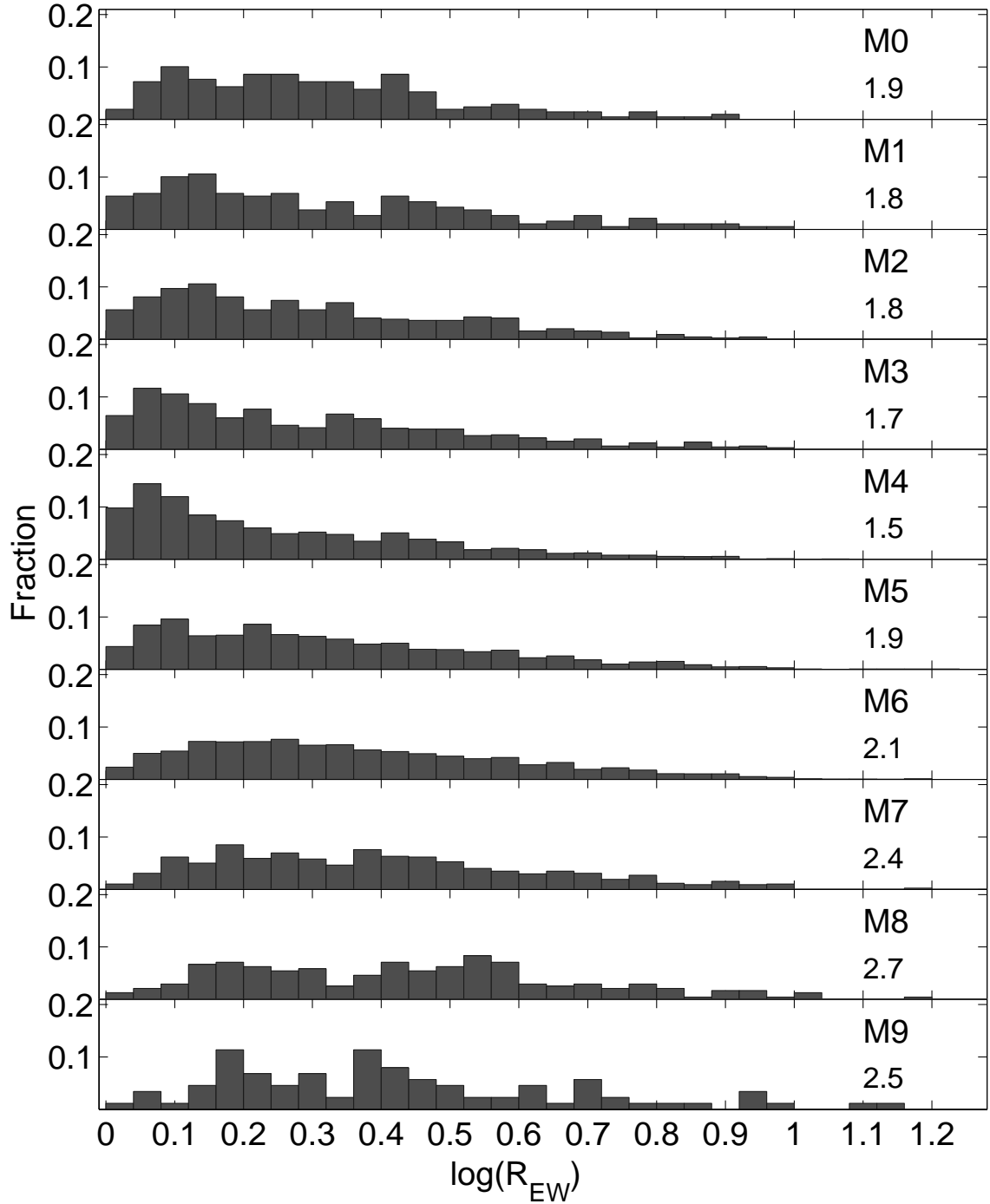,width=6.0in}}
\caption{Distributions of the ratio of maximum to minimum equivalent
width ($R_{\rm EW}$) for each spectral type.  The median $R_{\rm EW}$
value for each spectral type is listed on the right.  The overall
trend of increased variability with later spectral type is apparent.
\label{fig:maxminspectral}}
\end{figure}

\begin{figure}
\centerline{\psfig{file=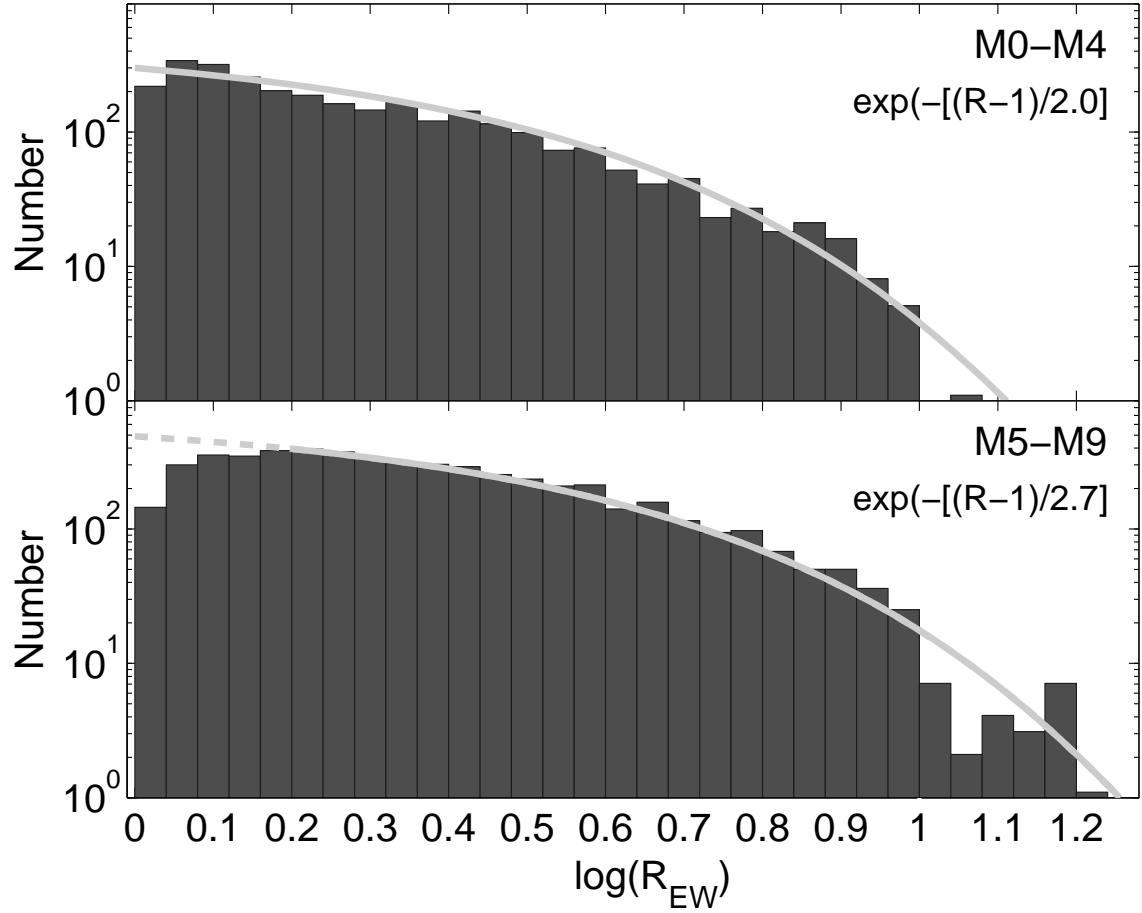,width=6.0in}}
\caption{Distributions of the ratio of maximum to minimum equivalent
width ($R_{\rm EW}$) binned for spectral types M0--M4 (top) and M5--M9
(bottom).  In both panels the gray line is an exponential fit with
resulting characteristic scales of $R_{\rm EW}-1\approx 2.0$ (M0--M4)
and $\approx 2.7$ (M5--M9).
\label{fig:rew_m04_m59}}
\end{figure}

The distribution of $R_{\rm EW}$ values for each spectral type is
shown in Figure~\ref{fig:maxminspectral}.  The more pronounced tail of
large $R_{\rm EW}$ values for spectral types $\gtrsim {\rm M4}$ is
evident.  In addition, we also find a systematic increase in the
median of the distribution, from about 1.8 for M0--M4 to about 2.5 for
M6--M9.  In Figure~\ref{fig:rew_m04_m59} we show the same
distributions, but combined for spectral types M0--M4 and M5--M9.  In
both cases the distributions are well fit by an exponential profile
with a characteristic scale of $R_{\rm EW}-1\approx 2$ for M0--M4 and
$\approx 2.7$ for M5--M9.  This indicates that high amplitude
variability is more common in late-M dwarfs.

\begin{figure}
\centerline{\psfig{file=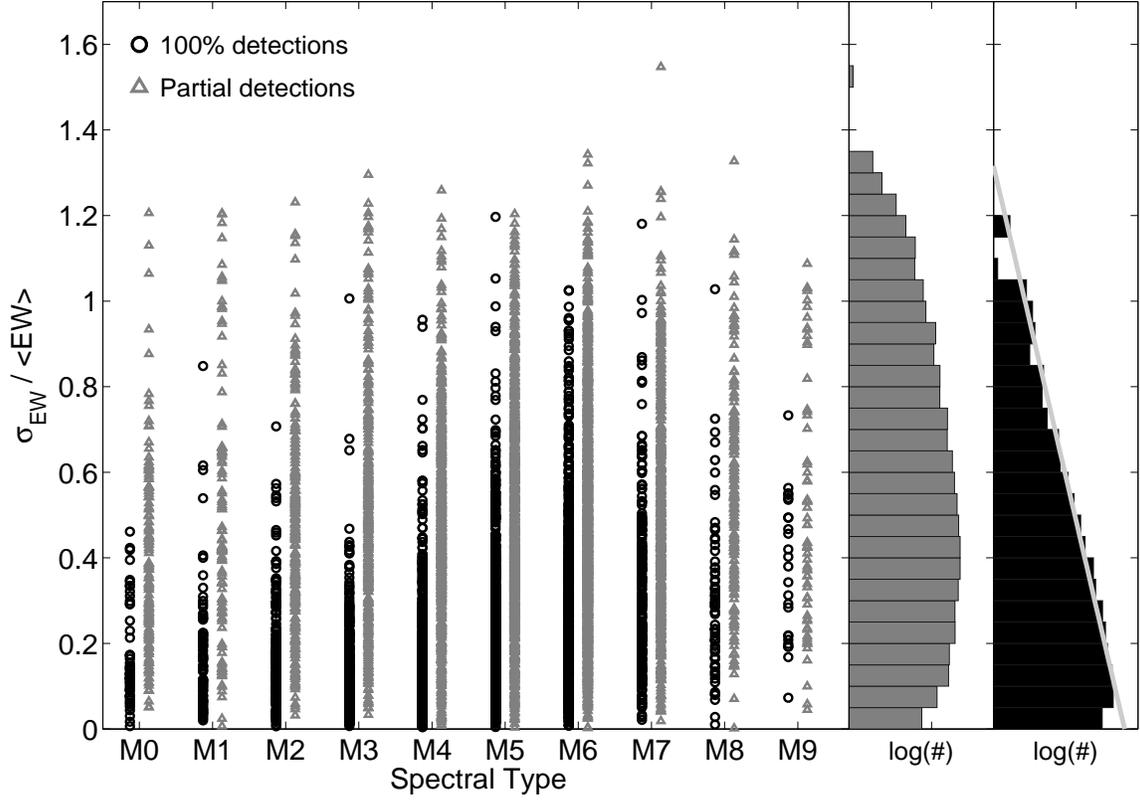,width=6.0in,angle=90}}
\caption{Ratio of the standard deviation to the mean equivalent width
for objects with detected H$\alpha$ emission as a function of spectral
type.  Circles designate objects classified as active in every
exposure, while triangles mark objects with at least one active
exposure and using upper limits on the equivalent width for the
inactive spectra.  The histograms for both samples are shown in the
right panels.  The light gray line is an exponential fit to the sample
of objects with $100\%$ detection fraction.
\label{fig:std}}
\end{figure}

\begin{figure}
\centerline{\psfig{file=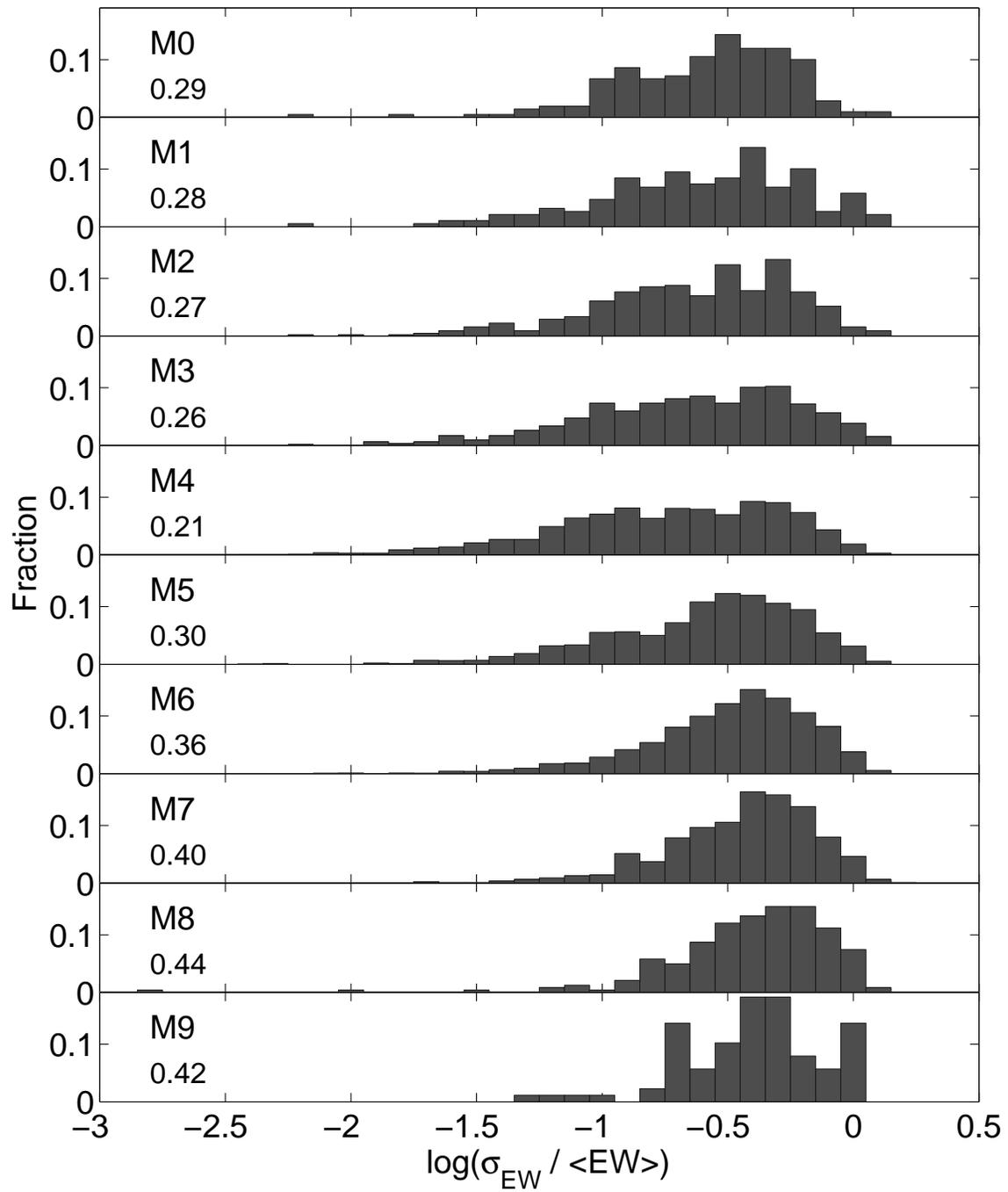,width=6.0in}}
\caption{Distributions of the ratio of the standard deviation to the
mean equivalent width for each spectral type.
\label{fig:stdspectral}}
\end{figure}

Finally, similar variability trends are found using
$\tilde{\sigma}_{\rm EW}$, with a steady increase from a maximum value
of about 0.4 for M0 to about 1 for M5--M7, followed by a decline to
about 0.7 for M8--M9 (Figure~\ref{fig:std}).  As in the case of
$R_{\rm EW}$, the distribution for partial detections is broader than
for objects with $100\%$ detections, which follows an exponential with
$N(\tilde{\sigma}_{\rm EW})\propto{\rm exp}(-\tilde{\sigma}_{\rm
EW}/0.18)$.  The distributions for each spectral sub-type are shown in
Figure~\ref{fig:stdspectral} and exhibit an increase in
$\tilde{\sigma}_{\rm EW}$ from a median value of about 0.28 for M0--M3
to about 0.4 for M6--M9.

\section{Discussion and Conclusions}
\label{sec:conc}

We study the variability of the H$\alpha$ chromospheric emission line
in M dwarfs using the individual 15 min spectroscopic exposures from
SDSS for an unprecedentedly large sample of 52,392 stars.  The typical
timescale probed by these data is 15 min to 1 hr.  About $16\%$ of the
sources exhibit H$\alpha$ emission in at least one exposure, and of
those about $45\%$ have H$\alpha$ emission in all of the available
exposures.  Only $26\%$ of all the objects with H$\alpha$ emission are
consistent with steady emission, spread relatively uniformly from M0
to M9.  The detection fraction as a function of spectral type exhibits
the known trend of a sharp increase from a few percent in M0--M3 to
tens of percent in later objects.  However, unlike the results of
previous studies based on SDSS pipeline-combined spectra and a fixed
equivalent width threshold of 1 \AA\ \citep{whw+04,whb+08}, we find a
peak detection fraction of $\sim 40\%$, with a steady decline beyond
M6.  We attribute this trend to our use of a spectral type dependent
detection threshold of $\sim 3-6$ \AA\ ($3\sigma$) in M5--M9 objects.

The H$\alpha$ variability, quantified in terms of $R_{\rm EW}$,
exhibits a substantial increase with later spectral type.  For M
dwarfs as a whole, the distribution of $R_{\rm EW}$ values appears to
follow an exponential with $N(R_{\rm EW})\propto{\rm exp}[-(R_{\rm
EW}-1)/2]$; the characteristic scale is about 2.7 for M5--M9.  We also
find that objects with partial detections exhibit a wider distribution
of $R_{\rm EW}$ values than those with persistent H$\alpha$ emission.
These results indicate that H$\alpha$ variability increases with later
spectral type, even as H$\alpha$ activity declines, and that M dwarfs
with intermittent H$\alpha$ activity are more variable than those with
persistent H$\alpha$ emission.

In the context of chromospheric heating, these trends suggest that the
magnetic energy input typically varies by a factor of about 2 on
timescales of $\sim 15$ min to $\sim 1$ hr.  Moreover, the increased
variability as a function of later spectral type, with a particularly
large increase beyond spectral types M3--M4, hints at an overall shift
in the magnetic field, leading to less uniform field dissipation.
Taken at face value, this conclusion suggests that the relative
contribution from small scales, which are likely to dissipate on short
timescales, increases for later M dwarfs.  This possible trend is in
conflict with preliminary results from Zeeman measurements of a few
objects in the range M0--M5, which point to an increase in the
relative contribution from large-scale fields
\citep{dmp+08,mdp+08,rb09}.  Alternatively, the increase in H$\alpha$
variability, particularly for sources with intermittent emission, may
be due to an increase in the stochastic dissipation of the field in
the presence of increasingly neutral atmospheres.  In this scenario,
heating of the bulk chromospheric plasma is suppressed by the
decoupling of the field from the atmosphere (leading to a decline in
$L_{\rm H\alpha}/L_{\rm bol}$), but small-scale stochastic heating
still takes place (leading to an increase in the H$\alpha$
variability).  The role of these two scenarios may become clearer as
samples of late-M dwarfs with Zeeman measurements become available.

\acknowledgements We thank Fergal Mullally and Steve Bickerton for
assistance with obtaining the individual SDSS spectra.  EAK
acknowledges financial support from the Harvard College Program for
Research in Science and Engineering (PRISE) and the Harvard College
Faculty Aide Program.  Funding for SDSS and for SDSS-II was provided 
by the Alfred P.
Sloan Foundation, the Participating Institutions, the National
Science Foundation, the U.S. Department of Energy, the National
Aeronautics and Space Administration, the Japanese Monbukagakusho,
the Max Planck Society, and the Higher Education Funding Council
for England.  The SDSS is managed by the Astrophysical Research
Consortium for the Participating Institutions. 


\clearpage

\clearpage

\clearpage

\clearpage

\clearpage

\clearpage

\clearpage

\clearpage

\clearpage

\clearpage

\clearpage

\clearpage

\clearpage

\clearpage

\clearpage

\end{document}